# Device Chemistry of Graphene Transistors

B. C. Worley,[1,3] S. Kim,[1,2] T.J. Ha,[4] S. Park,[1,2] R. Haws,[5] P. Rossky,[5] D. Akinwande,[1,2] and A. Dodabalapur[1,2]*

[1.] Microelectronics Research Center, The University of Texas at Austin, Austin, TX 78758, USA.

[2.] Department of Electrical and Computer Engineering, The University of Texas at Austin, Austin, TX 78712, USA.

[3.] Department of Chemistry, The University of Texas at Austin, Austin, TX, 78712, USA.

[4.] Department of Electronic Materials Engineering, Kwangwoon University, Seoul, Republic of Korea.

[5.] Department of Chemistry, Rice University, Houston, TX 77251, USA.

* Address correspondence to ananth.dodabalapur@engr.utexas.edu

**Abstract**. Graphene is an attractive material for microelectronics applications, given such favourable electrical characteristics as high mobility, high operating frequency, and good stability. If graphene is to be implemented in electronic devices on a mass scale, then it must be compatible with existing semiconductor industry fabrication processes. Unfortunately, such processing introduces defects and impurities to the graphene, which cause scattering of the charge carriers and changes in doping level. Scattering results in degradation of electrical performance, including lower mobility and Dirac point shifts. In this paper, we review methods by which to mitigate the effects of charged impurities and defects in graphene devices. Using capping layers such as fluoropolymers, statistically significant improvement of mobility, *on/off* ratio, and Dirac point voltage for graphene FETs have been demonstrated. These effects are also reversible and can be attributed to the presence of highly polar groups in these capping layers such as carbon-fluoride bonds in the fluoropolymer acting to electrostatically screen charged impurities and defects in or near the graphene. We also review the effects of other types of capping materials such as self-assembled monolayers and also gaseous species such as ammonia. In other experiments, graphene FETs were exposed to vapour-phase, polar, organic molecules in an ambient environment. This resulted in significant improvement to electrical characteristics, and the magnitude of improvement to the Dirac point scaled with the dipole moment of the delivered molecule type. This type of experimental data is supported by recent theoretical work, wherein the interactions of polar molecules with impurities such as charged ions or adsorbed water on a graphene surface were simulated. The potential profile produced in the plane of the graphene sheet by the impurities was calculated to be significantly reduced by the presence of polar molecules. We present strong evidence that the polar nature of capping layers or polar vapour molecules introduced to the surface of a graphene FET act to mitigate detrimental effects of charged impurities/defects.

## 1. Introduction

Graphene is an attractive material for a variety of applications, most notably, those in the realm of microelectronics. High mobility, carrier velocity, and operating frequency, atomic-layer thickness, and good stability are among the numerous characteristics which have been well studied both theoretically and experimentally after the isolation of graphene by Geim and Novoselov in 2004 (1-10). These characteristics make graphene very favourable for use in field-effect transistors (FETs) and radio frequency (RF) devices (11-25). However, much work needs to be done with regard to both scaling up graphene production and integration into existing semiconductor manufacturing processes (18, 26, 27). Such processing incorporates defects and impurities in graphene, which are charged, resulting in a degradation of electrical characteristics (10, 16, 28-48). Various groups have reported methods that have been successful at mitigating the effects of these impurities and defects (12-17, 40-42, 45, 46, 49-60). In depth study of the chemical processes behind the mitigation of the effects of charged impurities and defects on graphene are important, so that this material may be better employed in microelectronics. Recent computational studies focused on two of most prevalent categories of impurities (61). Additionally, graphene-based materials such as graphene nanoribbons should also be studied both because of their technological promise and because they possess substantially higher concentrations of charged defects and impurities.

Two-dimensional materials such as graphene are, in many ways, ideal for the study of impurities and defects since these can be easily accessed. In comparison, the defects/impurities are mostly buried and difficult to access in bulk three-dimensional materials. This aspect of the problem allows a more clear demonstration of the effects of many reversible chemical treatments on graphene device behaviour. In turn, demonstration of these effects enables a clearer understanding of impurity/defect-related phenomena. Other groups have also reported on similar experiments.

This review will describe the effects of impurities and defects on charge transport and device performance. The utility of polar molecules – both in thin-film form and in vapour phase – in reducing the impact of impurities and defects will be demonstrated. We will describe both experiments with films of material containing polar molecular groups, and the impact of annealing and reordering on such films. Next, we will discuss experiments with polar vapours, including the similarities and advantages over thin-film experiments. Finally, we will briefly describe theoretical work in support of experimental results.

## 2. Graphene Background

### 2.1. Transport Physics

The foundations of graphene electronic theory were established by Geim and Novoselov. Important properties of graphene physics include its zero-band gap, extraordinarily high charge carrier mobility values, and ballistic transport over sub-micrometer length scales (1, 2, 4, 8, 62). Building on this knowledge, Popov and others, found that graphene is uniquely aromatic when considered in finite molecular sizes. Its π-electrons are pairwise localized to each hexagon ring (63). Zubarev and co-workers expounded on this idea, describing the delocalization of graphene π-electrons as having greater complexity than traditional descriptions of aromaticity, where a finite graphene sheet develops "finite-size domains of perturbed bonds" in response to a perturbation of the π-electron system (64). Nevertheless, graphene's π-electrons can best be described as delocalized, resulting in excellent electrical conductivity and high mobilities. It is also important to note that graphene's electrons behave as Dirac fermions, having both zero effective mass and mean free paths on the order of one micron (1, 2, 9, 29, 65).

These qualities of graphene charge carriers result in very high, though substrate- and interface-dependent, mobility and carrier velocity values, some of which even approach values characteristic of ballistic transport

(65, 66). High mobility values result in excellent conductivity for graphene, which also has a linear dependence on carrier density (4, 6, 32). Conduction in graphene is also favoured by the unique band structure, where the conduction and valence bands meet at the charge-neutral Dirac point (44, 67). However, graphene's unique electron-hole puddles have been cited as the cause of lower than expected mobility values for graphene samples, which is indeed just one of many factors contributing to limiting the mobility of charge carriers in graphene. Yet, it is unlikely that the puddles are the dominant, and certainly not the only source of charge carrier scattering in graphene samples (68). Therefore, the remainder of our review will concern the most likely culprits of charge scattering: charged impurities and defects.

*2.2. Charged Impurities and Defects*

We have mentioned many important and favourable qualities of graphene transport physics above. However, given the vigorous and ongoing debates over exact details of graphene transport in theoretical versus experimental results, low temperature versus high temperature, suspended versus on-substrate, and other specific cases, we must qualify the particular situation under which we will discuss graphene in this review. We focus here on experimental graphene that was produced via a chemical vapour deposition (CVD) growth process on copper substrate (60, 69). The graphene is transferred onto $SiO_2$/Si substrates using a wet transfer process (40, 41, 60, 69-72). This method is one of the most commonly used in the fabrication of graphene based FET devices, and has seen much improvement in recent years (60, 73-75). The types of devices discussed here have exposed graphene, which is open to access by capping layers or vapours. Thus, contamination of the sample by charged impurities and defects such as adsorbed or trapped oxygen and water, residue from the transfer, or trapped ions and substrate defects are significant contributions to disorder (28, 33, 38, 50, 51, 53, 57, 76-85). Specifically, scattering of charge carriers is an important mechanism by which charged impurities and defects degrade electronic transport in graphene.

The fine structure constant α, as described by Das Sarma and co-workers, provides a good reference point from which to understand how it might be possible to reduce charge scattering in graphene (35). Equation 1 illustrates how increasing the dielectric constant values ($\kappa_x$) for the environments above and below graphene can reduce α.

$$\alpha = \frac{2e^2}{(\kappa_1 + \kappa_2)\hbar v_F} \qquad [1]$$

Das Sarma and co-workers reported that a smaller α value results in increased Coulombic scattering- limited mobility. Their work, and similar work by other groups, offers insight into how to improve graphene's electrical properties. Polar molecules can help in two possible ways. The first way involves partial neutralization of charged impurities by polar molecules, and the second way involves the alignment of the polar molecules into a dipole layer, which increases the κ around graphene, decreasing α to produce the improvements mentioned above (35, 86, 87). Such alignment is driven by the total energy of the system which would be minimized in this configuration, and is aided by the polarizability of graphene (88-91).

While a variety of other scattering sources, such as phonons and surface corrugations, do affect the conductivity ($\sigma$) of graphene, scattering by charged impurities (*CI*) affects the conductivity, dependent on the charge carrier density($n$), according to the Equation 2 described by Fuhrer and co-workers (32, 33).

$$\sigma_{CI}(n) = C_{CI} e \left| \frac{n}{n_{imp}} \right| \qquad [2]$$

$C_{CI}$ is a constant of value $5 \times 10^{15}$ V$^{-1}$s$^{-1}$. Their experimental results with potassium ions and graphene FETs showed both reduced mobility with increasing impurity concentration. Their results also supported theoretical predictions that the conductivity has a linear relation to gate voltage when charged impurity scattering is the dominant scattering mechanism. Charge carriers in graphene undergo ballistic transport until deflected either by charged impurities (long-range scattering) or by one another (short-range scattering). More specifically, charged impurities and defects act to scatter charge carriers in graphene by influencing the two mechanisms of Coulombic, long-range scattering and short-range scattering. Coulombic, long-range scattering shortens the mean free path for charge carriers, and this is observed in lower-than-predicted mobility values. Charged impurities and defects may also cause shifts in the Dirac point (charge-neutral point) voltage of measured current versus gate voltage curves ($I_d$-$V_g$) (28, 30, 42, 92-94). Dirac voltage shifts are problematic both in that they represent non-ideal behaviour of graphene and in that a device with a significantly shifted Dirac point voltage would be more difficult or less efficient to operate as a normal transistor. Hysteresis in such $I_d$-$V_g$ curves is another observable, unfavourable electrical characteristic that indicates charge scattering or changes in charge carrier density due to charged impurities in the form of adsorbed water or other polar molecules (28, 38, 78, 84).

## 3. Mitigation of Charged Impurity Scattering

### 3.1. Fluoropolymer Coatings

Having recognized the challenges which graphene's susceptibility to charged impurities and defects presented to the future of graphene microelectronics, work on new methods of improving key performance metrics in graphene devices is of great importance. Such methods would need to interface with existing microelectronics industry processes and equipment. Covalent modification chemistry was not considered for both this reason and because such functionalization may suppress graphene charge carrier mobility by interrupting the sp$^2$ hybridized bond structure of graphene (51). Instead, Ha *et al.* used vacuum sublimation to deposit semiconducting organic capping layers onto a bottom-gate, graphene FET (13). Thin films of both α-sexithiophene (α-6T) and hexadecafluorocopperphthallocyanine (F$_{16}$CuPC) were employed as capping layers atop the graphene. The inset of Figure 1 shows the basic chemical structure of F$_{16}$CuPC. While both organic materials improved the measured electrical characteristics of the graphene device in air, the authors noted that the fluorinated organic material, F$_{16}$CuPC, had particularly favourable effects. The authors observed significant improvements to the on/off current ratio, Dirac voltage point, and field-effect mobility of the graphene device treated with the F$_{16}$CuPC. These improvements to electrical characteristics stand in stark contrast with other prior attempts to place additional layers atop a graphene device. Figure 1 shows the results of testing a graphene FET with F$_{16}$CuPC capping layer in air. The fluorinated organic material significantly improved the *on/off* current ratio, a key metric for transistor performance in terms of high speed and low leakage current. The Dirac peak voltage, where resistance is a maximum and conductivity is a minimum, is also shown in Figure 1 as significantly positive when measured on the initial bare graphene device. This indicates some residual doping of the graphene by impurities and defects. However, upon capping with $F_{16}$CuPC, the Dirac peak shifts back toward the ideal graphene Dirac voltage of zero gate voltage. Ideally, the Dirac voltage should be at zero volts and the minimum conductivity (at the Dirac point) should be as small as possible. The diffusive transport model used to extract values for both field-effect mobility and residual carrier density at the minimum conductivity point is also plotted in Figure 1, and gives a good fit to the experimental data. In addition to the improvements to electrical characteristics listed above, the authors also calculated that $F_{16}$CuPC both significantly increased the mobility and reduced the residual carrier density at the minimum conductivity point.

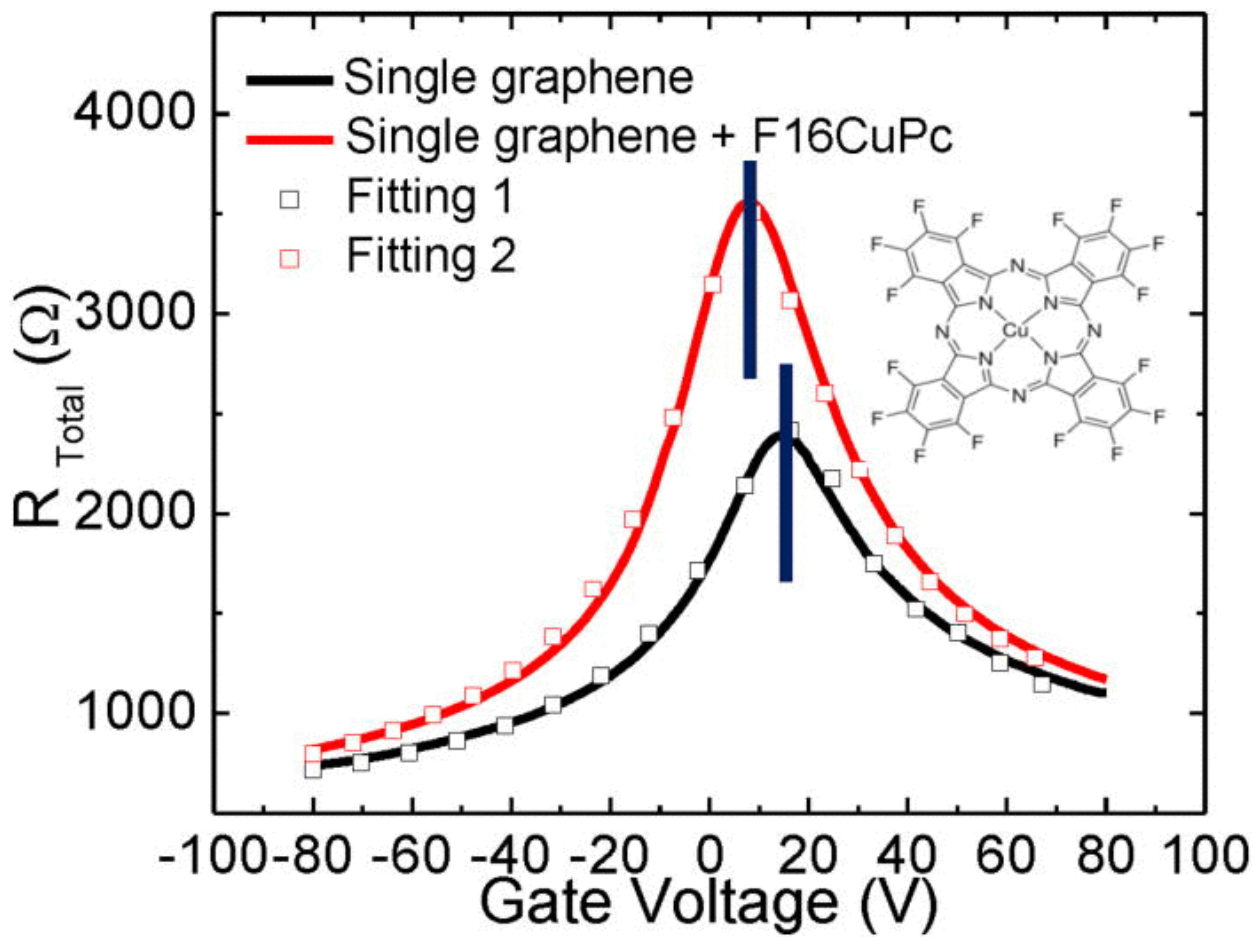

Figure 1. Total resistance as a function of gate voltage for a graphene FET before and after capping with $F_{16}$CuPC, diffusive transport model fitting, and, inset, the basic chemical structure of $F_{16}$CuPC.

Ha and co-workers' work with $F_{16}$CuPC led to further study the effect of other fluorinated materials on graphene, as described in work published in References (16, 40, 41). The authors chose to employ the fluoropolymers CYTOP® and Teflon-AF. The fluoropolymers were deposited as a thin film directly on top of the graphene layer of a FET via spin-coating methods, with subsequent annealing. Figure 2 shows the effect of a spin-coated thin film of the CYTOP® on the transfer characteristics of a graphene FET, with improvements to electrical characteristics similar to those previously observed with a capping layer of the fluorinated organic molecule $F_{16}$CuPC. The CYTOP® fluoropolymer capping layer significantly reduced the off-state current at the Dirac voltage point (red curve), seen as a large increase in maximum measured resistance compared to that of the bare graphene device (black curve). Remarkably, after removal of the CYTOP® layer, the measured characteristics of the graphene device appear to return toward their initial values (blue curve, Figure 2a). In the experiments, the authors did not achieve complete removal of the fluoropolymer from the graphene, which may account for the incomplete reversion of the characteristics. More importantly, this degradative reversion in electrical characteristics indicates that the impact of the CYTOP® layer upon graphene is caused by a reversible interaction. The *on/off* ratio is critical for graphene FET operation, and significant improvement in the ratio is shown more clearly as a sharpening of the graphene+ CYTOP® peak versus the peak for graphene alone in Figure 2b, a plot of normalized resistance versus gate voltage. Additionally, the improvements caused by CYTOP® are much better than those caused by capping with pentacene, and sharply contrast the degradative effects of a $SiO_2$ capping layer.

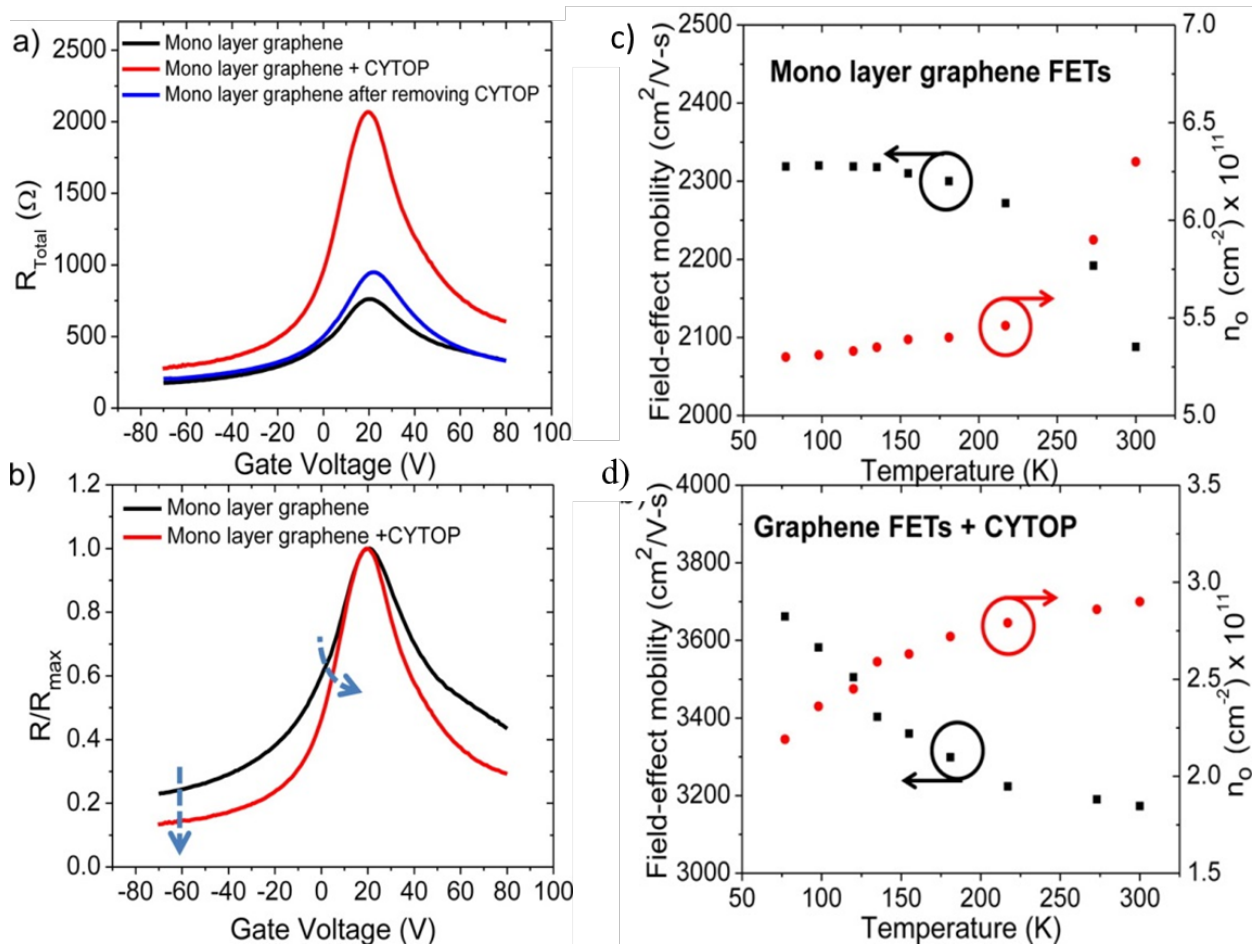

Figure 2. a) Reversible effects of a fluoropolymer film on the total resistance as a function of gate voltage for a graphene FET. b) Effects of a fluoropolymer film on the normalized resistance as a function of gate voltage, showing improved *on/off* ratio. c) Mobility and residual carrier concentration as a function of temperature for a graphene FET. d) Improved mobility and residual carrier concentration as a function of temperature for a fluoropolymer-capped graphene FET. Reprinted with permission from (40). Copyright 2013 American Chemical Society.

CYTOP® also caused great improvement in graphene's temperature-dependent charge carrier mobility. In comparing Figure 2d (CYTOP®-coated graphene FET) with Figure 2c (bare graphene FET), note that the scattering-limited mobility with CYTOP®-coating is higher than that of the bare graphene device at all measured points. Further, the mobility in Figure 2d is shown to continuously increase as temperature decreases, instead of plateauing as in Figure 2c. Concurrent with improvements to mobility, CYTOP® caused significant temperature-dependent reduction in residual carrier concentration, also shown in Figures 2c and 2d. Ha *et al.* attributed these improvements to a reduction in short range scattering, which lowers the value of the residual carrier concentration (35). Residual carrier concentration is a metric related to the concentration of impurities in or around the graphene. Additionally, the authors observed that increased annealing temperatures gave greater impact of the fluoropolymer to device characteristics, and they attributed that effect to a reorganization of the C-F bond dipoles atop the graphene. Finally, the apparent return of the pre-fluoropolymer electrical characteristics of graphene further indicate that the actions of the fluoropolymer upon graphene are of a noncovalent, reversible nature (40).

Continuing the work of employing polar fluoropolymers to improve graphene device electrical characteristics, Ha *et al.* observed the significant favourable shift in the Dirac voltage peak for graphene FET devices upon capping and annealing with CYTOP® and Teflon-AF fluoropolymer layers shown in Figures 3a & b. Before fluoropolymer deposition, the bare graphene devices exhibited a very positively-shifted Dirac voltage peak position, due to doping effects from charged impurities and defects which incorporate in and around the graphene from both the wet transfer process and the substrate. Asymmetry between electron and hole transport is also evident. Upon coating with CYTOP® and gradual annealing in nitrogen atmosphere from 30°C to 180°C, or coating with Teflon-AF and gradual annealing in nitrogen atmosphere from 30°C to 300°C, the authors observed dramatic, favourable shifts in the Dirac voltage peak position from the very positively-shifted initial state toward nearly zero gate voltage. Having now observed such movement of the Dirac voltage peak by several different fluorinated organic molecule/polymer capping layers, the authors attributed the favourable Dirac voltage shifts to mitigation by the polar C-F bonds of the molecules/polymers of the Coulombic charge scattering effects of charged impurities and defects around graphene. Electron and hole transport became more symmetric. Again, they reported improvement in mobility concurrent with decrease in residual carrier concentration, and Figure 3c shows significant improvement to the *on/off* ratio for fluoropolymer-coated graphene FETs.

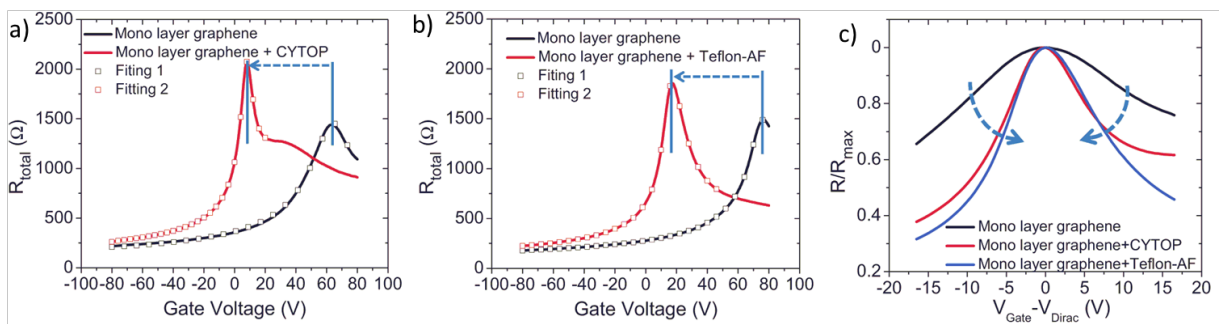

Figure 3. Favourable shift in Dirac voltage peak for graphene FETs upon application of a) CYTOP® and b) Teflon-AF fluoropolymers. c) Improvements to *on/off* ratio for graphene FET upon coating with fluoropolymers. Reprinted with permission from (41). Copyright 2013 IEEE.

In further studies with fluoropolymers on graphene devices, Ha *et al.* observed changes to a greater variety of devices that indicate the favourable effects of polar, C-F bonded materials are of a widely applicable nature (16). As shown in Figures 4a and 4b, treatment with fluoropolymers can significantly improve the Dirac peak of a graphene FET, regardless of p- or n-doping from charged impurities and defects (red curves). The authors attributed the adaptability of the fluoropolymer's interactions with either p-type or n-type charged impurities and defects to self-organization of the dipolar C-F bonds upon heat treatment (annealing). Again, the improvements to the device electrical characteristics exhibit reversible behaviour upon removal of the fluoropolymer (blue curves), indicating a noncovalent, reversible interaction. They also reported the highest *on/off* ratio at room temperature for CVD-grown graphene at the time, indicated in Figure 4c. The greatest improvements to electrical characteristics were observed for devices where the graphene was capped on both sides by fluoropolymer further. This observation supports the assessment that charged impurities and surface defects of the substrate are significant causes of scattering for charge carriers in graphene. The results also support the assessment that the polar nature of fluoropolymers mitigates that scattering to improve charge transport in graphene, which manifests as measurable improvements to electrical characteristics. Additionally, the authors observed that the hydrophobic nature of the fluoropolymers used in the experiments serves to protect the capped graphene devices from water, another common contaminant known to frequently degrade device characteristics (16, 28, 38, 78, 84).

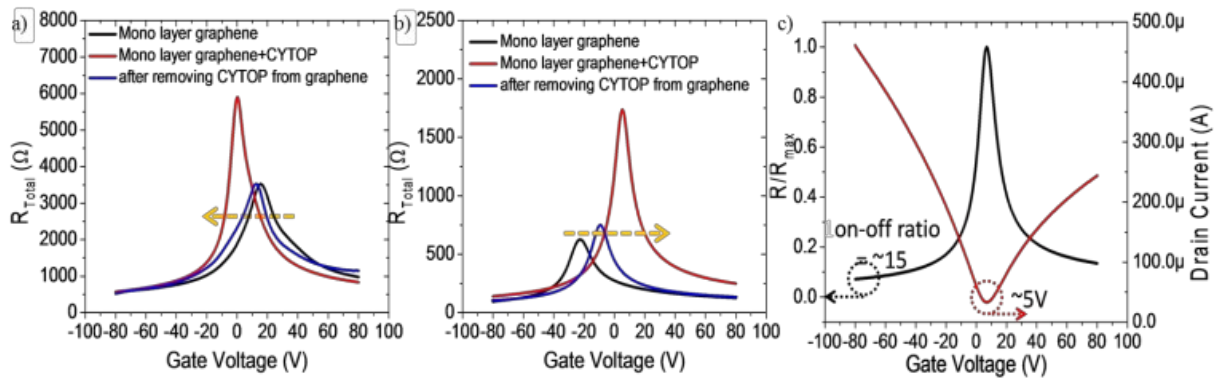

Figure 4. Reversible effects of fluoropolymer in favourably shifting the Dirac voltage for both a) p-doped and b) n-doped graphene FET. c) Improved electrical characteristics of fluoropolymer-capped graphene FET include *on/off* ratio and Dirac peak voltage. Reprinted with permission from (16). Copyright 2013 IEEE.

It is also important to note that, as shown in Figure 5, Ha *et al.* have observed these improvements for 30 different graphene FET samples (41). While these devices were fabricated at different times in different batches, they all exhibit improved mobility and favourable Dirac voltage shifts toward zero gate voltage upon capping with the fluoropolymer CYTOP®. Such statistical analysis is important because measured results can quantitatively vary from device to device, so it is best to check for repeatable results over a large number of devices. Figure 5 shows that, while the quantitative results do vary from sample to sample, the qualitative effects of the fluoropolymer on the graphene FETs' electrical characteristics are uniform.

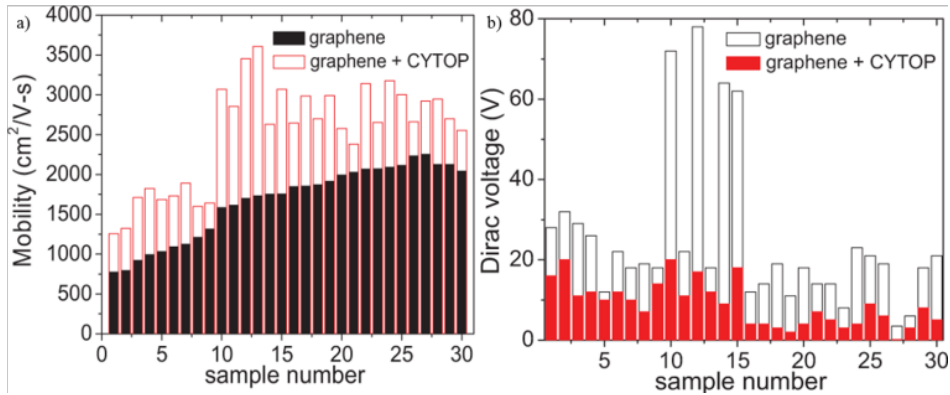

Figure 5. a) Mobility values and b) Dirac peak voltages before and after fluoropolymer coating for 30 different graphene FET samples. Reprinted with permission from (41). Copyright 2013 IEEE.

In addition to the work on graphene, Ha *et al.* also observed improvements to other 2-D materials by treatment with fluoropolymers, such as increased drain current, mobility, and transconductance in reduced graphene oxide, and improved *on/off* current ratio in $MoS_2$ (16). Kim and co-workers reported significantly increased air-stability over time with black phosphorous (95). These results with fluoropolymers and a variety of 2-D materials hold great promise for this method to be both practical and impactful in developing the full potential of graphene and other 2-D materials in the microelectronics industry.

### 3.2. SAMs and Other Coatings

Other groups have achieved improvements to graphene device characteristics via coating methods such as treatment with various self-assembled monolayers (SAMs). Upon coating of graphene alone or of both graphene and underlying substrate with HMDS, and comparison with graphene-on-$SiO_2$ device characteristics, Akinwande and co-workers found significant improvement to graphene device mobility values, reduction in residual carrier concentration, and favourable shifts in Dirac voltage peak, similar to results with fluoropolymers (46). The authors also attributed the observed improvements to the polar nature of the HMDS molecules acting to both enhance the dielectric screening and mitigation of charged impurities via a reduction of the dimensionless fine structure constant. It is also important to note that the hydrophobic methyl groups of the HMDS molecules which contact the graphene separate or block the graphene from any adsorbed impurities on the substrate or in the air. In their studies on alkyl phosphonic acid-based self-assembled monolayers, Cernetic and co-workers reported improvement to graphene device mobility, favourable shifts in Dirac voltage peak, and reduction in hysteresis (53). They partly attributed these improvements to reduction of defects/charge trap states in the layer immediately underneath graphene. Many other groups using SAMs report similar improvement to graphene device electrical characteristics, where the dipole moments of a variety of SAM molecules cause shifts in the Dirac voltage peak position and improve the mobility relative to those of graphene-on-$SiO_2$ devices (50, 51, 54, 56, 79, 96, 97) All of these groups also acknowledge the significant detriment to graphene transport caused by scattering from charged impurities and defects at the substrate.

### 3.3. Polar Vapour Molecules

Ha *et al.* observed from fluorinated organic molecule and fluoropolymer treatment of graphene devices that it is possible to drastically improve electrical characteristics such as mobility, on/off current ratio, Dirac voltage peak, and residual carrier concentration. They learned that these improvements arise from the mitigation or neutralization of charge scattering from charged impurities and defects in and around graphene via interaction of the polar C-F bonds of the fluorinated organic molecules and fluoropolymers with charged impurities and defects. The authors found that heat treatment allows reorientation of the polar bonds of the fluoropolymers to better mitigate charged impurities and defects. Ruoff and co-workers found that residual PMMA on graphene devices can cause unfavourable p-doping of the graphene (45). Instead of depositing thin films or SAMs, they used a liquid soak treatment with polar formamide molecules to counteract PMMA doping with favourable n-doping. After formamide treatment, they reported favourable Dirac voltage shifts toward zero gate voltage and increased mobility values. The authors also found that, upon vacuum treatment, the favourable effects of the formamide molecules disappeared as the molecules evaporated from the devices, indicating a reversible behaviour. Further testing of how other types of polar molecules might affect graphene in a similar fashion could be accomplished using simpler experimental methods through the employment of polar vapour molecules. Polar small molecules such as ethanol, with well-defined dipole moments, are excellent model systems to help study the effect of such materials on graphene properties. The experimental advantages of using vapour-phase polar molecules as opposed to applying thin films of polar molecules/polymers include easy application, easier removal in the form of simple evaporation in ambient, and in situ reorientation of the vapour molecules' dipoles around charged impurities and defects (as opposed to annealing of thin films). As early as 2007, Schedin and co-workers were able to detect a variety of polar vapour molecules adsorbed on graphene (98). They measured changes in resistivity of a graphene device upon separate adsorptions of ammonia ($NH_3$), carbon monoxide (CO), water ($H_2O$), and nitrogen dioxide ($NO_2$) molecules. $NH_3$ and CO caused positive changes to resistivity, while $H_2O$, and $NO_2$ caused negative changes. The authors attributed these positive and negative valued changes to electron and hole doping, respectively. Theoretical work by Leenaerts *et al.*(82), in which they computationally studied adsorption of the same molecules on graphene, supported the experimental results of Schedin *et al.* (98). Both groups concluded that $H_2O$, and $NO_2$ accept transfer of electronic charge from graphene, while $NH_3$ and CO donate charge to graphene. Leenaerts *et al.* quantified the adsorption energies for each polar molecule-graphene interaction as below 0.1 eV. According to Umadevi and Sastry, who also computationally studied interaction between polar small molecules and graphene-like structures, such interaction energies as those reported by Leenaerts *et al.* for $NH_3$, CO, $H_2O$, and $NO_2$ fall in the range best described as physisorption (88). In such cases of physisorption between vapour-phase molecules and a solid surface like graphene, there is little perturbation of the electronic structures (88, 99, 100). Umadevi and Sastry also reported that metal atoms have stronger interactions energies with graphene than do polar small molecules, and that the respective metal ions interact with even greater strength toward graphene.

To replicate similar results with polar vapour molecules and graphene FETs as seen with the methods discussed above, Worley *et al.* exposed graphene FETs on an open-air probe station at room temperature to vapours of various polar molecules (42). Experimental details can be found in Reference (42). Figure 6 shows the results for their experiments with exposure of graphene FETs to polar vapours of acetone, ethanol, and isopropyl alcohol (IPA). The effects of the polar vapours on the Dirac peak voltage of a graphene FET are shown in Figure 6a. Nitrogen, the carrier gas, has little to no effect on the Dirac peak position, as compared to its p-doped (positively shifted) state as measured on the device in ambient air before vapour treatment. However, upon exposure of the device to polar vapours, there is both significant improvement to the *on/off* current ratio and significant favourable movement of the measured Dirac peak position toward zero gate voltage for each type of vapour employed. Figure 6b quantitatively shows the relationship between the theoretical dipole moment, a measure of the average dipole present in each molecule based on the separation of its positively- and negatively-charged regions or atoms, of each type of vapour and the magnitude by which it favourably shifts the Dirac peak position of the graphene FET. This behaviour was observed in several samples fabricated at different times, and is illustrated in Fig. 6b in which data from 3-4 samples is averaged and plotted. In addition, Figure 6c shows the improvements to both hole and electron mobility as a function of theoretical vapour molecule dipole moment. The accompanying reduction in residual carrier concentration is shown in Figure 6d as a function of theoretical vapour molecule dipole moment.

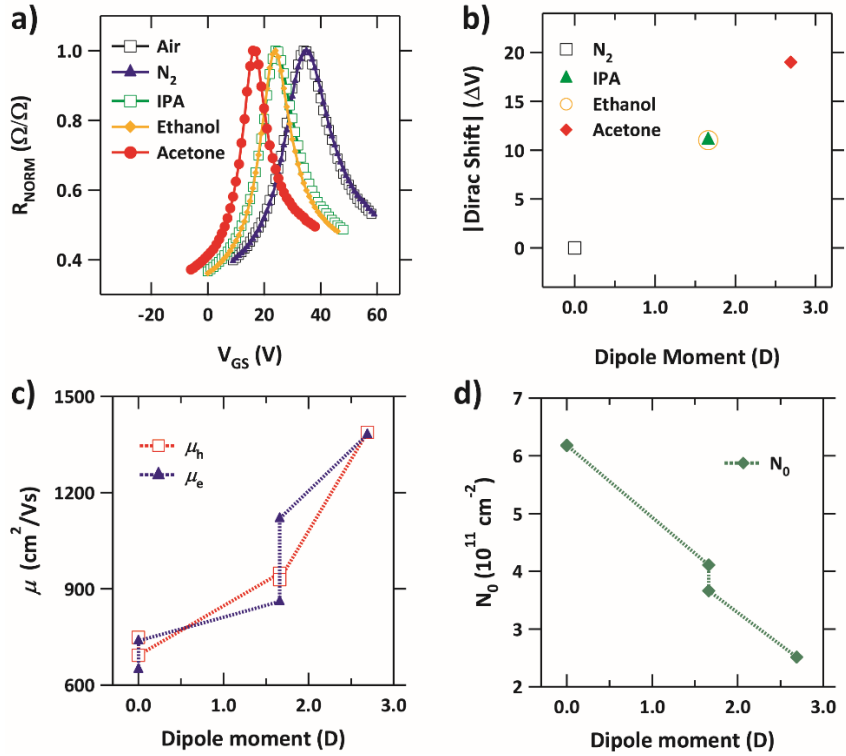

Figure 6. a) Favourable shifts in Dirac voltage for a graphene device exposed to various polar vapours. b) Magnitude of Dirac voltage shift, c) charge carrier mobility values increases, and d) residual carrier concentration reductions correspond to dipole moment of impingent vapour.

Figure 7 reveals that vapours of hexane, a nonpolar molecule, exhibit little to no effects on the Dirac peak position of the graphene FET, while acetone causes a significant favourable shift of the Dirac voltage peak on the same FET. The data supports assertions that the improvements to graphene electrical characteristics do indeed depend on the polarity of the impingent vapour molecules. To test for reversibility in the effects of polar molecules on graphene (like Ha *et al.* observed with fluoropolymer treatments), Worley *et al.* also tested the graphene FET periodically after initial exposure to acetone vapour.

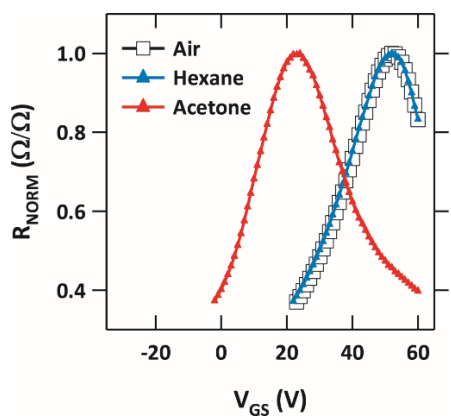

Figure 7. Nonpolar hexane exhibits almost no effect upon electrical characteristics of graphene FET, but polar acetone exhibits a large change in the Dirac voltage peak.

Figure 8a shows the Dirac voltage peaks for a graphene FET as measured initially (under nitrogen), upon exposure to acetone vapour, and then at periodic measurements over time after the delivery of vapour was stopped. After stopping the delivery of acetone vapour, the reversion of the Dirac peak position toward the initial state of the device is evident. Worley *et al.* attributed this reversion to gradual desorption of the acetone vapour molecules from the graphene surface in ambient over time. As the polar molecules desorb and leave the graphene, they cease to mitigate charge scattering by charged impurities and defects. Consequently, the magnitude of improvements to Dirac peak (Figure 8b), mobility (8c), and residual carrier concentration (8d) caused by the delivered polar acetone vapours all decrease. Thus, with polar vapour molecules, the authors observed the same effects exhibited on a graphene FET as Ha *et al.* did with fluoropolymer treatment. Worley *et al.* also observed that these effects are reversible, indicating a noncovalent electronic interaction. They further assessed that the nature of this interaction is electrostatic, where the polar molecules act to screen or neutralize charge carrier scattering by charged impurities and defects on graphene.

Similar results for vapour-phase modification of graphene were recently reported by Ago and co-workers, who used both vapour-deposition and spin-coating methods to adsorb polar piperidine molecules on the surface of graphene devices (101). For both methods, the authors reported significant favourable shifts in the Dirac voltage peak and increased mobility. They also attributed the interaction between piperidine and graphene to be a noncovalent one, just as Worley *et al.* concluded in their own experiments.

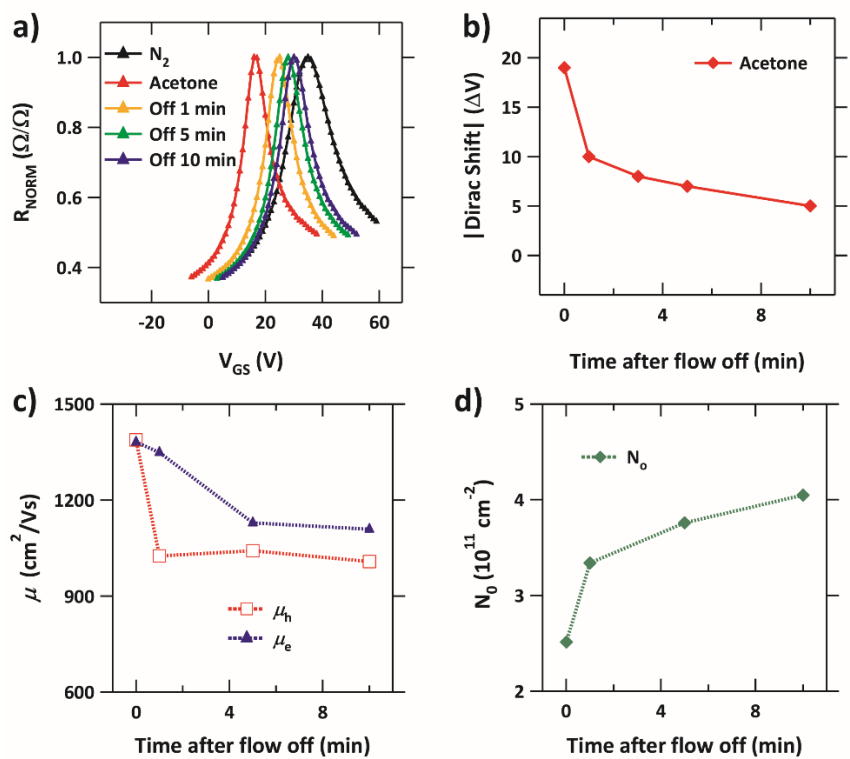

Figure 8. a) Initial Dirac voltage shift with acetone exposure, and subsequent reversal over time. b) Magnitude of Dirac voltage shift, c) charge carrier mobility values decreases, and d) residual carrier concentration increases correspond to dissipation of acetone molecules with time.

## 4. Theoretical Understanding of Charged Impurity Effect Mitigation

To better understand how polar molecules act to mitigate the effects of charged impurities and defects around graphene, Worley *et al.* performed computational chemistry simulations (61). These consisted of prototypical charged impurities, specifically, a sodium ion and a water molecule, interacting with a graphene sheet. There have been many theoretical studies, both quantum mechanical and using molecular dynamics, on various atoms, ions, and molecules interacting with graphene (77, 81, 82, 87, 88, 93, 99, 102-122). In particular, Ao *et al.* reported that, in their density-functional theory simulations, carbon monoxide more strongly chemisorbs to Al-doped graphene than to pristine graphene (99). Worley *et al.* studied such interactions both with and without the presence of various quantities of acetone, IPA, and ethanol molecules, which were chosen to match the polar molecules used in their vapour-phase experiments (61). The authors used molecular dynamics software to perform the simulations, and calculated the electrostatic potential caused by impurities in the plane of the graphene sheet, both with and without polar molecules present. Further detail on their theoretical methods can be found in Ref. (61).

### 4.1. Results with Sodium Impurity

Figure 9 illustrates a snapshot of a model system under study. Here, a graphene sheet, a sodium ion, and fifty adsorbed acetone molecules comprise the simulation system, where the sodium ion is a good example of any given point charge type of impurity common to graphene devices (33, 88, 99, 103, 116, 119-124).

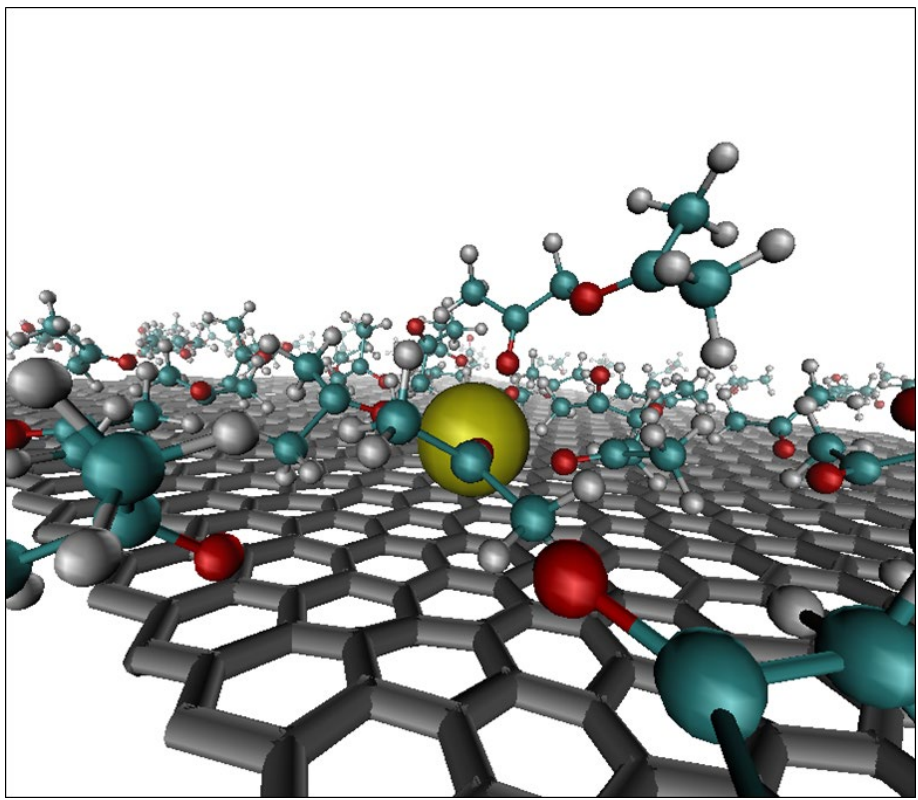

Figure 9. Snapshot of a simulation involving acetone molecules solvating a sodium ion atop a graphene sheet.

After 10 nanoseconds of NVT (constant number of particles, volume, and temperature) statistical ensemble simulation at room temperature, the authors were able to examine the trajectories and atomic charges of the molecules/atoms to calculate the radial electrostatic potential out to 1 nm from a point on the graphene plane directly beneath the sodium ion. Figure 10a shows a "slice" of the radial potential plot. The magnitude of potential in the plane of the graphene directly under the sodium ion in the absence of polar molecules is large (black curve), and rapidly decays over distance away from the ion. However, various polar molecules solvated the sodium ion and significantly reduced the potential at the graphene sheet (red, green, and blue curves). The inset of Figure 10a shows that the magnitudes of reduction in potential by polar molecules scale with their respective calculated dipole moments (Figure 10b-d). Acetone, having the greatest dipole moment of the three types of molecules (Figure 10d), most significantly reduced the charged sodium ion potential profile in the plane of the graphene sheet. This calculated trend is in good agreement with experimental vapour-phase results, where improvements to graphene device electrical characteristics scale with respective theoretical dipole moments of polar vapour molecules delivered to the surface of a graphene FET.

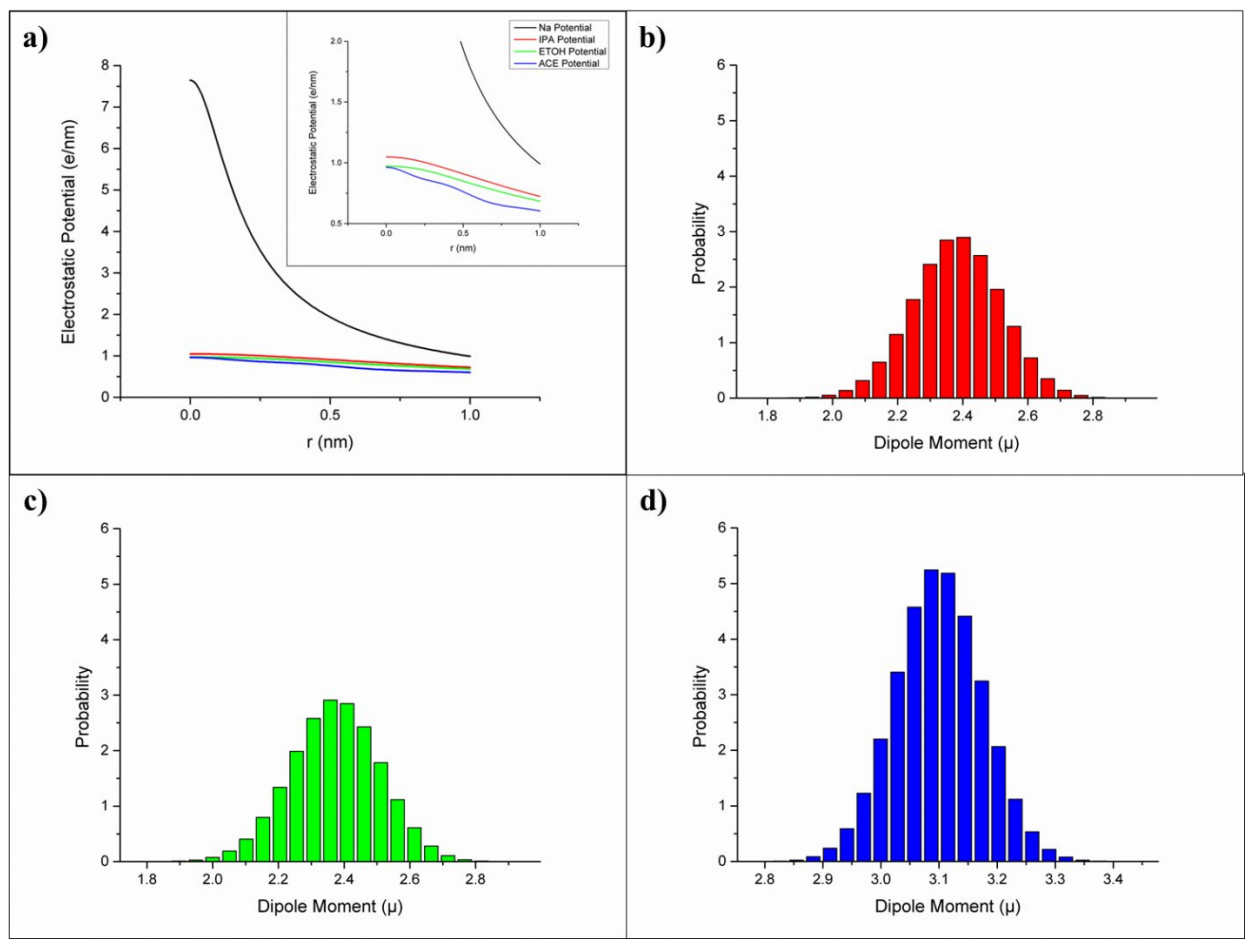

Figure 10. a) Electrostatic potential in the plane of the graphene sheet from the sodium ion as a function of distance with and without the effects of various polar molecules. Probability distributions for dipole moments of b) IPA, c) ethanol, and d) acetone molecules calculated from their respective atoms' positions and point charges in simulations.

Another major observation made from their calculations is the identification of two mechanisms by which polar molecules act to mitigate the sodium ion impurity potential profile in the plane of the graphene sheet. The authors calculated the magnitude of change in the potential profile from that of a graphene and sodium ion system alone to that after inclusion of acetone molecules. This magnitude of change in potential is represented by the black curve in Figure 11a. The red curve represents the contribution to this potential change due to one mechanism, physical displacement of the sodium away from the graphene sheet by acetone molecules. This creates a so-called solvent-separated ion-graphene structure. The blue curve represents the contribution due to electrostatic screening of the sodium ion charge by orientation of the acetone dipoles around the ion, the second mechanism. Although these two mechanisms are distinct effects, they are not completely separable, as the screening would be quantitatively affected by the ion position. Figure 11b shows that, at short range from the point on the graphene sheet below the sodium ion, displacement is the dominant mechanism contributing to the potential magnitude reduction. At greater distance, screening becomes the dominant mechanism.

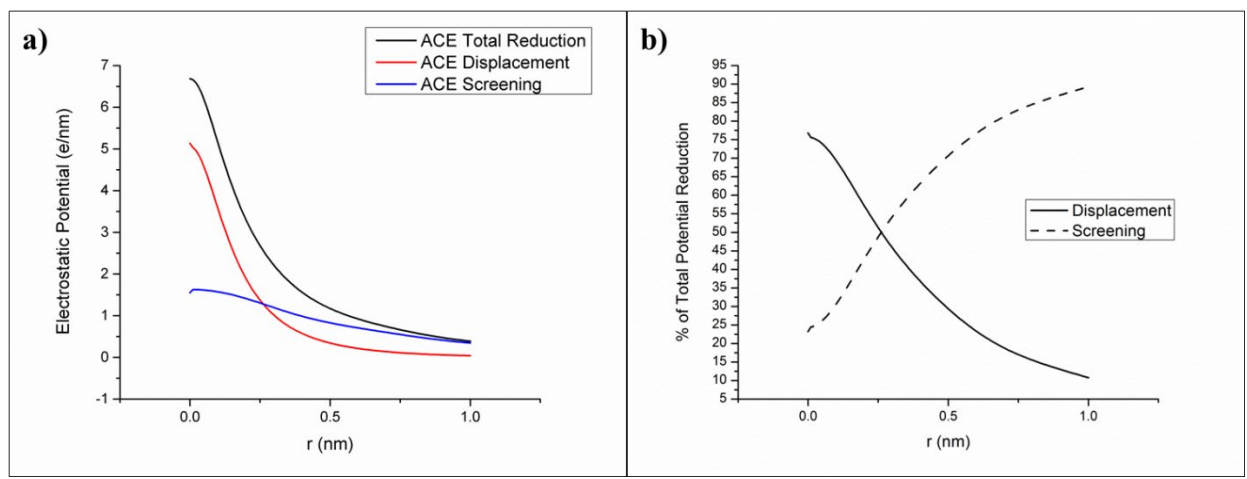

Figure 11. a) Potential plot showing the magnitude of potential magnitude reduction for an adsorbed water molecule impurity by acetone molecules as well as contribution of both displacement and of screening mechanisms to this potential reduction. b) Percentage contribution of each mechanism to the potential reduction.

Worley *et al.* also studied the effects of varying the number of acetone molecules in the graphene/sodium system. Figure 12a shows that the impurity potential profile magnitude varied with the number of acetone molecules present, and that greater numbers of acetone molecules more significantly reduced the potential profile. As displacement of the sodium ion away from graphene by acetone molecules is a key mechanism of potential reduction, Figure 12b shows that greater numbers of acetone molecules also further displaced the sodium ion. To further detail how different numbers of acetone molecules displace an impurity like sodium from graphene, the authors plotted the relative probability of the sodium ion being found at a given distance from the graphene sheet for each time step of the simulations with different numbers of acetone molecules, as shown in Figure 12c. Numbers of acetone molecules as low as ten were sufficient to form a solvation shell around the sodium ion, with a distribution centered at about 3.5 Angstroms above the graphene sheet. As more acetone molecules are included in the simulation box, the distribution became bimodal, which represented the formation of a second solvation shell around sodium. Similar results were obtained with other polar molecules. The authors concluded from these observations that the majority of sodium impurity screening can be achieved with relatively low numbers of adsorbed polar molecules.

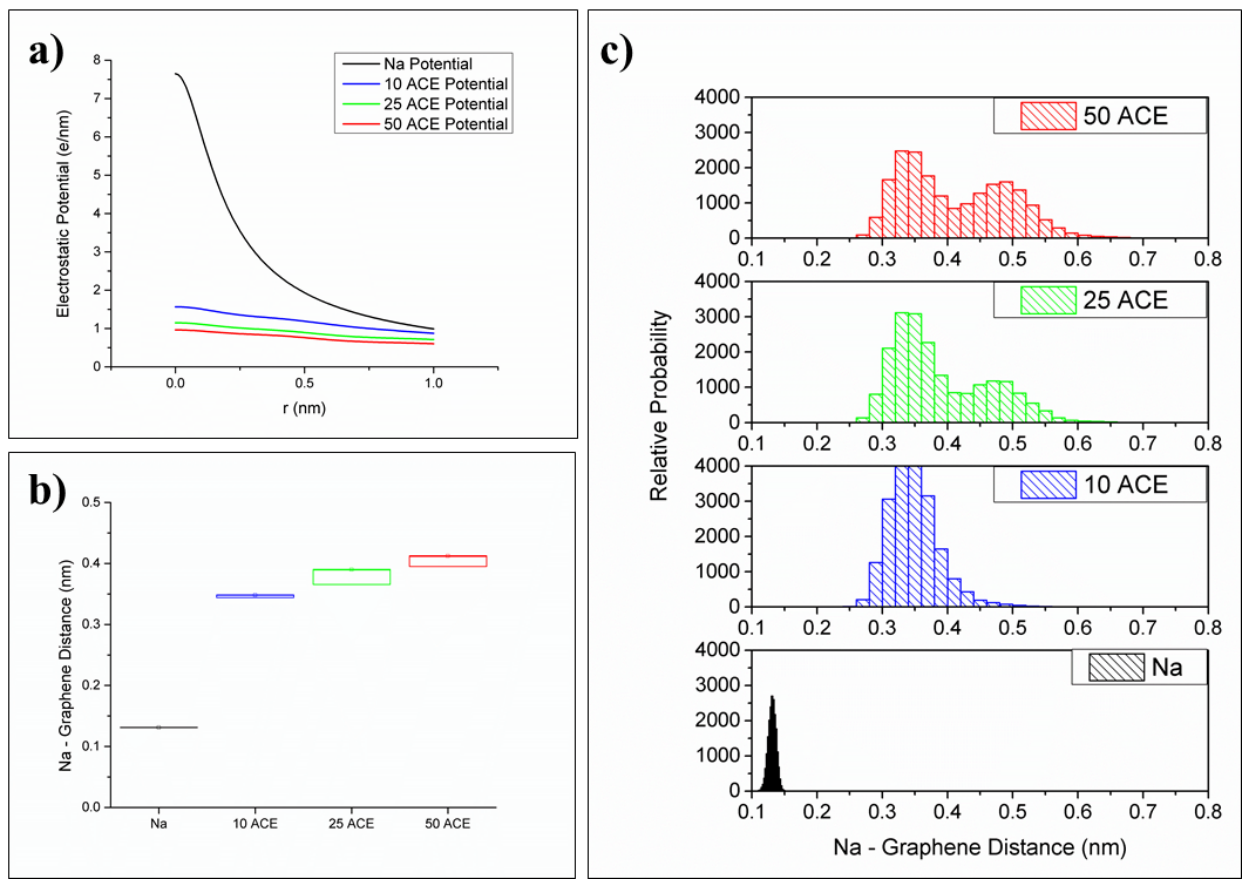

Figure 12. a) Plots of potential profile for sodium, both alone and with various numbers of acetone molecules. b) Mean displacement (small square) of sodium ion away from graphene as a function of number of acetone molecules, with standard error (larger box boundaries) boxes. c) Distribution of values of sodium ion displacement by acetone molecules, revealing formation of a second solvation shell with greater numbers of acetone molecules.

*4.2. Results with Water Impurity*

In addition to the point charge type impurity like sodium, Worley *et al.* also chose to simulate water as a representative molecular dipole type of impurity on graphene. Water is a ubiquitous impurity that impacts electrical characteristics of many experimentally-studied graphene devices (28, 38, 78, 84), and which has also been studied theoretically (81, 82, 87, 112, 113, 117). Figure 13a shows a one-dimensional representation of the potential profile in the plane of the graphene sheet caused by a single water molecule oriented such that one hydrogen faces down toward the sheet [orientation details in Ref. (61)]. The authors calculated significant reduction from the potential profile of water alone atop graphene (black curve) to much lower potentials upon incorporation of acetone (blue curve), IPA (red curve), and ethanol molecules (green curve). The polar molecules electrostatically induce a reorientation of the dipolar water molecule atop the sheet, which causes the potential in the plane of the sheet to change from positive to negative potential. (The complexities involved in this process are discussed in full in the computational paper (61).) Regardless of the sign of the potential, the magnitude of potential caused by a water impurity in the plane of the graphene sheet is significantly reduced by polar molecules, as shown in Figure 13b, where the absolute values of the curves in Figure 13a are plotted. As was the case with a sodium ion impurity, this data strongly supports the hypothesis that the polar nature of acetone, ethanol, and other molecules acts to reduce the electrostatic potential of charged impurities on graphene, which serves to mitigate the charge scattering effects of such impurities. Yokota *et al.* similarly found that, without altering the electronic structure of graphene, the dipole moments of polar SAM molecules alter the electrostatic potential which affects a graphene layer (50).

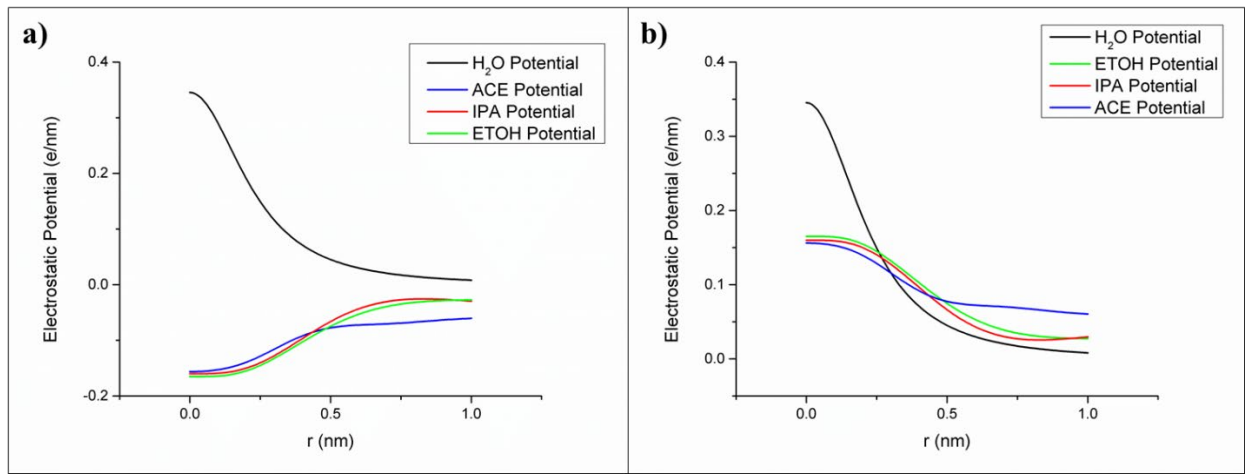

Figure 13. a) Electrostatic potential in the plane of the graphene sheet from the water molecule as a function of distance with and without the effects of various polar molecules. b) Absolute value of potential profiles show in part a.

## 5. Conclusions

This review focused on the mitigation of impurities and defects in as-fabricated graphene devices. If graphene is to play a successful role in the future of microelectronics, charge scattering from impurities and defects inherent to manufacturing and processing must be reduced. Only then can the highly favourable electrical properties of graphene be fully realized. Using polar coatings of fluoropolymers, polar molecules, or SAMs, it is possible to greatly improve graphene device characteristics such as mobility, on/off ratio, and Dirac point voltage. Both the polar nature of these polymers/molecular structures and the reversible, noncovalent nature of their interactions served to greatly mitigate charge scattering. These coating experimental results lead to experimentally simpler methods using polar vapours, where physisorption is the likely interaction. With these studies, it was found that the magnitude of improvements to key graphene device characteristics strongly correlated with the dipole moments of exposed polar vapours. These improvements were observed to be reversible upon desorption of the polar molecules from the graphene surface. The authors hypothesized that the improvements to graphene transport are due to electrostatic, noncovalent interactions between polar molecules and both charged impurities and defects in and around the graphene. Computational studies indicated that polar molecules can indeed act to reduce the electrostatic potential in graphene that is caused by charged impurities. The magnitude of potential reduction by each type of polar molecule scaled well with the calculated dipole moment of the respective polar molecule. These theoretical results lend strong support to the hypothesis regarding improvements to graphene transport. Studying these interactions has led to better understanding of how to improve the quality of electronic transport in graphene devices. That these improvements can be applied to other 2-D systems is encouraging for future efforts.

**Acknowledgements**

We gratefully acknowledge support from NSF ECCS Division, NSF EFRI Division Grant **EFMA-1542747**, and NSF Cooperative Agreement No. EEC-1160494 (NASCENT). PJR gratefully acknowledges the support of NSF Chemistry Division (CHE-1362381). We also acknowledge Rod Ruoff for pioneering work in CVD growth of graphene.